\documentclass[aps,prapplied,twocolumn,showpacs,superscriptaddress]{revtex4-2}
\usepackage{graphicx}
\usepackage{amsmath}
\usepackage{amssymb}
\usepackage{tabularx}
\usepackage{float}
\usepackage{color}
\usepackage[normalem]{ulem}

\newcommand{\e}{\textup{e}}
\newcommand{\ii}{\textup{i}}
\newcommand{\red}[1]{{\color{red}{#1}}}

\graphicspath{{converted_graphics/}}

\begin{document}
\title{Discrimination of vortex and pseudovortex beams with a triangular optical cavity}

\author{L. Marques Fagundes}
\affiliation{Departamento de F\'isica, Universidade Federal de Santa Catarina, Florian\'opolis, SC, 88040-900, Brazil}

\author{P. H. Souto Ribeiro} %\email[]{p.h.s.ribeiro@ufsc.br}
\affiliation{Departamento de F\'isica, Universidade Federal de Santa Catarina, Florian\'opolis, SC, 88040-900, Brazil}

\author{R. Medeiros de Ara\'ujo}
\affiliation{Departamento de F\'isica, Universidade Federal de Santa Catarina, Florian\'opolis, SC, 88040-900, Brazil}
\email{renne.araujo@ufsc.br}

\date{\today}
\begin{abstract}
A triangular optical cavity can be used to distinguish between two beams with the same intensity profile but different wavefronts. This is what we show in this paper, both theoretically and experimentally, in the case of beams with a doughnut-like intensity profile: one of them having a helical wavefront (vortex beam with orbital angular momentum) and the other with no orbital angular momentum at all (which we call pseudovortex beam). We write the mode decomposition of such beams in the Hermite-Gaussian basis and in the Laguerre-Gauss basis, respectively, and study how they interact with a triangular cavity in terms of their resonance peaks. The experimental results corroborate the theoretical predictions, demonstrating that each beam exhibits a distinct resonance pattern. This suggests that such a cavity can be used to identify beams carrying orbital angular momentum, effectively distinguishing them from pseudovortices. Moreover, we propose an experiment where three cavities may be used to filter out the pseudovortex from a superposition of vortex and pseudovortex.

\end{abstract}
\maketitle
%
%%%%%%%%%%%%%%%%%%%%%%%%%%%%%%%%%%%%%%%%%%%%%%%
%  SECTION
%%%%%%%%%%%%%%%%%%%%%%%%%%%%%%%%%%%%%%%%%%%%%%%
\section{Introduction}

Vortex beams, characterized by their helical phase fronts and orbital angular momentum (OAM), offer unique properties that can be harnessed for applications such as high-capacity optical communication \cite{gbur2007vortex, liu2019high, wang2024optimization, zhu2017free, kai2017orbital, wang2016advances, ndagano2018, yuan2022}, quantum information processing \cite{zhang2022parallel, vaziri2002experimental, tempone2016optical, vicuna2014classical,vallone2014}, and precision microscopy \cite{verbeeck2010production, tian2015resolution, plociniczak2016transformation, tao2005fractional, wang2017}. Over the past three decades, the development of an optical toolbox for manipulating structured light, particularly vortex beams, has enabled increasingly precise control over the generation, transformation, and detection of these beams, facilitating novel experimental capabilities and enhancing the versatility of structured light in practical applications.

Devices such as spatial light modulators (SLMs) and q-plates have been extensively utilized to create OAM beams. SLMs modulate the phase of light through a programmable interface, enabling the generation of custom wavefronts with specific topological charges \cite{rosales2017shape}. Meanwhile, q-plates utilize anisotropic birefringence to convert circularly polarized light into beams with quantized OAM states in a compact and efficient way \cite{khonina1992phase,karimi09}.

Manipulating and reading OAM beams have also seen significant advancements. Devices such as refractive and diffractive elements, including spiral phase plates, and forked diffraction gratings, allow for the sorting and multiplexing of OAM modes \cite{Kotlyar2017,Volyar2019,kai2017orbital, yan2014high,Wen2020,Soifer1994,Khonina2004, Hickmann2010,Mourka2011,Anderson2012,Berkhout2010,Mirhosseini2013,Zhou2017}, enabling the use of multiple channels in optical communication systems. Other techniques including optical autocorrelation measurements \cite{Kumar2022} or mode intensity analysis \cite{Volyar2020a,Volyar2020b} have also been successfully employed to identify and sort such optical modes.

%Optical correlation to read $\ell$ \cite{Kumar2022}.

%DOE \cite{Soifer1994,Khonina2004}, triangular aperture \cite{Hickmann2010,Mourka2011,Anderson2012}, intensity method \cite{Volyar2020a,Volyar2020b}, cylindrical lens \cite{Kotlyar2017,Volyar2019}.

Recent innovations like metasurfaces have further enhanced the control over OAM beams by integrating subwavelength structures that manipulate the light field with high precision \cite{liu2021multifunctional, qin2018transmission, ji2019dual, zhang2023metasurface}.

A device of particular relevance in the present study is the optical cavity. Previous work has demonstrated that a linear optical cavity can effectively discriminate the OAM content of a light beam \cite{Shibiao2020}. Specifically, such a cavity is capable of distinguishing between Laguerre-Gaussian (LG) modes with different topological charges $\ell$, provided that the radial indices $p$ are the same. Additionally, recent research conducted by our group has shown that a triangular optical cavity can discriminate between Hermite-Gaussian (HG) modes \cite{Santos2021}. This ability arises from the fact that the triangular cavity has an odd number of mirrors, which breaks parity symmetry and enables the distinction between symmetric and antisymmetric modes (e.g., $HG_{01}$ versus $HG_{10}$).

In the present work, we demonstrate that it is possible to detect the presence of OAM even with a triangular cavity by analyzing the distribution of resonance peaks. To substantiate this, we compare the resonance peaks of a Laguerre-Gaussian (vortex) beam with those of a ``pseudovortex" beam, defined here as a beam that exhibits the intensity profile of an LG mode but has null topogical charge (meaning it has a flat wavefront). This analysis positions the triangular optical cavity as a viable tool for identifying OAM, due to its inherent interferometric properties. Our scheme can be further developed to outperform other systems in the ability of filtering out pseudovortex modes from vortex modes (independently of their topological charge).

%%%%%%%%%%%%%%%%%%%%%%%%%%%%%%%%%%%%%%%%%%%%%%%
%  SECTION
%%%%%%%%%%%%%%%%%%%%%%%%%%%%%%%%%%%%%%%%%%%%%%%
\section{Vortex versus pseudovortex beams}
\label{sec:symmetry}

We begin this section by analyzing the mode decompositions of the vortex and pseudovortex beams, using two natural cavity-resonant bases: the Laguerre-Gauss and Hermite-Gauss modes.

The vortex beam is described in the Laguerre-Gauss basis by a single mode, $LG_{10}(r,\phi,z)$ ($\ell=1$ and $p=0$), where $r$ and $\phi$ are the transverse polar coordinates and $z$ is the longitudinal coordinate. At the focal plane $z=0$, the transverse profile of the (normalized) vortex beam reads
\begin{equation}
    V(r,\phi) = LG_{10}(r,\phi,0) = \frac{2r}{\sqrt{\pi}\,w^2_0}\e^{-r^2/w_0^2}\e^{\ii\phi}\,,
    \label{eq:LG10}
\end{equation}
where $w_0$ is the beam waist and the exponential $\e^{\ii\phi}$ accounts for the beam's orbital angular momentum. Changing from polar $(r,\phi)$ to cartesian coordinates $(x,y)$ \red{immediately reveals} the vortex's HG-decomposition. Since $r\e^{\ii\phi}=x+\ii y$ and $r^2=x^2+y^2$, we get:
\begin{equation}
    V(r,\phi) = \frac{1}{\sqrt{2}}\left[\,HG_{10}(x,y,0)+\ii\,HG_{01}(x,y,0)\,\right]\,.
    \label{eq:V-decomposition}
\end{equation}

In this work, the \textit{pseudovortex beam} is defined as the beam that has the same doughnut-like intensity profile as the vortex beam presented above, but differs from the latter by its plane wavefront, losing the azimuthally dependent term:
\begin{equation}
    PV(r,\phi) = LG_{10}(r,\phi,0)\e^{-\ii\phi} = \frac{2r}{\sqrt{\pi}\,w^2_0}\e^{-r^2/w_0^2}\,,
    \label{eq:LG10PV}
\end{equation}
where $PV$ stands for the pseudovortex transverse profile. 

Although we consider here, for simplicity, the case $\ell=1$ and $p=0$, other vortex-pseudovortex pairs may also be defined for different values of $\ell$ and $p$. This is done in a way that the pseudovortex profile intensity matches that of a $LG_{\ell p}$ mode, as discussed in Section \ref{sec:discussion}.

We wish to write the pseudovortex of Eq.~(\ref{eq:LG10PV}) as a linear combination of Laguerre-Gaussian profiles. In order to do that, we need to calculate the coefficients $a_{\ell p}$ of the decomposition
\begin{equation}
    PV(r,\phi) = \sum_{\ell=-\infty}^{+\infty} \sum_{p=0}^\infty\, a_{\ell p} LG_{\ell p}(r,\phi,0)\,,
\end{equation}
with
\begin{equation}    
    a_{\ell p} = \int_0^\infty rdr \int_0^{2\pi} d\phi\ PV(r,\phi)LG^*_{\ell p}(r,\phi,0)\,,
    \label{eq:a_lp}
\end{equation}
\vspace{-5pt}
\begin{equation}
    LG_{\ell p}(r,\phi,0) = \frac{A_{\ell p}}{w_0}\left(\frac{r\sqrt{2}}{w_0}\right)^{|\ell|}
    \!\!L^{|\ell|}_p\!\left(\frac{2r^2}{w_0^2}\right)\e^{-\tfrac{r^2}{w_0^2}}\e^{\ii\ell\phi}\,,
    \label{eq:LG_lp}
\end{equation}
where $A_{\ell p}$ is a normalization factor and $L^\alpha_p(x)$ is the generalized Laguerre polynomials.

It is easy to see that an analytical calculation of the integral (\ref{eq:a_lp}) gives zero for any $\ell\neq 0$: since $PV(r,\phi)$ actually does not depend on $\phi$, the azimuthal integral simplifies to $\int_0^{2\pi} d\phi\,\e^{-\ii\ell\phi}$, which is equal to $2\pi\delta_{\ell,0}$. Thus, the cylindrical symmetry of the pseudovortex's \textit{field} ensures that no mode with orbital angular momentum participates in its decomposition. From a complete calculation (see Appendix), we obtain:
\begin{equation}
    PV = \frac{\sqrt{\pi}}{2}LG_{00}-\frac{\sqrt{\pi}}{4}LG_{01}-\frac{\sqrt{\pi}}{16}LG_{02}-...
    \label{eq:PV-decomposition}
\end{equation}
where the dependency on $(r,\phi)$ has been omitted for compacity. Interestingly, only even-order modes participate in this decomposition ($N=|\ell|+2p$), which, again, is related to the cylindrical symmetry of the pseudovortex.

It is also possible to decompose this pseudovortex in the Hermite-Gauss basis: it suffices to write each term $LG_{0p}$ in the HG basis (to an analytical basis-changing formula, see for example \cite{Beijersbergen1993}). However, the decomposition into the LG basis, besides being simpler, is also more suitable for comparison with the experimental data. In fact, each mode $LG_{0p}$ displays a single resonance peak, whereas OAM modes $LG_{\ell 0}$ are split into two components (symmetric and antisymmetric), as will become clear in Sections \ref{sec:results} and \ref{sec:discussion}.

\begin{figure}[H]
   \includegraphics[width=\columnwidth]{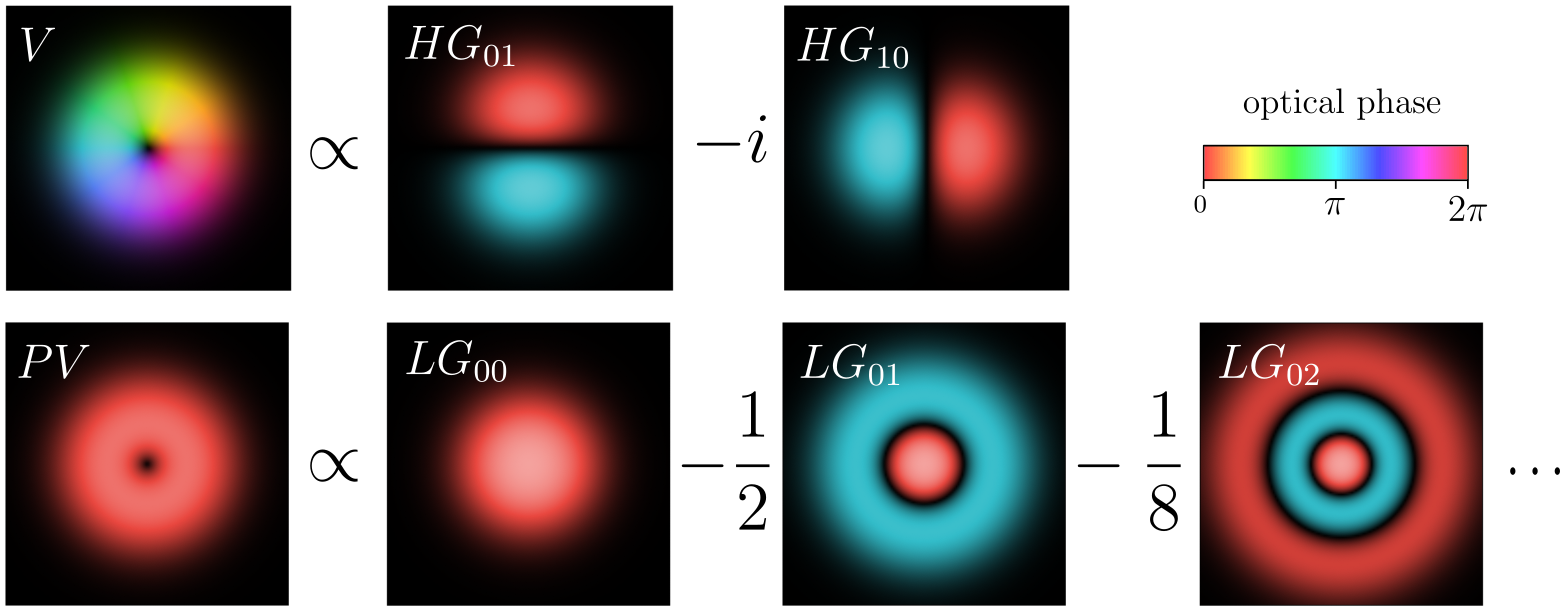}
   \caption{Mode decomposition of the first-order vortex and pseudovortex beams depicted in amplitude-phase profiles. The amplitude and phase are encoded, respectively, into brightness and hue. The intensity contribution of each component is proportional to the squared modulus of its coefficient.}
   \label{fig:vortex-pseudovortex}
\end{figure}

Let us now examine Eqs.~(\ref{eq:V-decomposition}) and (\ref{eq:PV-decomposition}) to analyze how each mode contributes to the intensity of the light fields. It is clear that the vortex beam ($LG_{10}$) gets all of its intensity from first-order modes (in the HG basis, we have one half from the $HG_{10}$ mode and the other half from $HG_{01}$), whereas the pseudovortex has no energy at all on the first order. In fact, $\approx$ 78,5\% of its energy emanates from the zero-order gaussian beam and other $\approx$ 19,6\% originates in second-order modes ($LG_{01}$ or, equivalently, $HG_{20}$ and $HG_{02}$), leaving less than 2\% for higher-order modes. Figure \ref{fig:vortex-pseudovortex}) provides a graphical representation of the mode decomposition of the vortex and pseudovortex modes (Eqs. \ref{eq:V-decomposition} and \ref{eq:PV-decomposition}).

%%%%%%%%%%%%%%%%%%%%%%%%%%%%%%%%%%%%%%%%%%%%%%%
% FIG
%%%%%%%%%%%%%%%%%%%%%%%%%%%%%%%%%%%%%%%%%%%%%%%
%%%%%%%%%%%%%%%%%%%%%%%%%%%%%%%%%%%%%%%%%%%%%%%%

The take-home message of this section is that, although both beams share the same \textit{intensity} profile at the waist plane, their contrasting phase profiles play a crucial role in their spatial mode decomposition, feature that we intend to capture with an optical cavity, as described in the following section. 

It is worth mentioning that the contrasting phase profiles also affect the propagation of the beams. While the vortex beam maintains its shape as it diverges, the intensity profile of the pseudovortex beam loses the central hole in the far field, resembling a Gaussian profile (the apparent vortex is erased, which is why we call it a pseudovortex).

\section{Experimental Setup}
\label{sec:exp}

\begin{figure}[H]
   \includegraphics[width=\columnwidth]{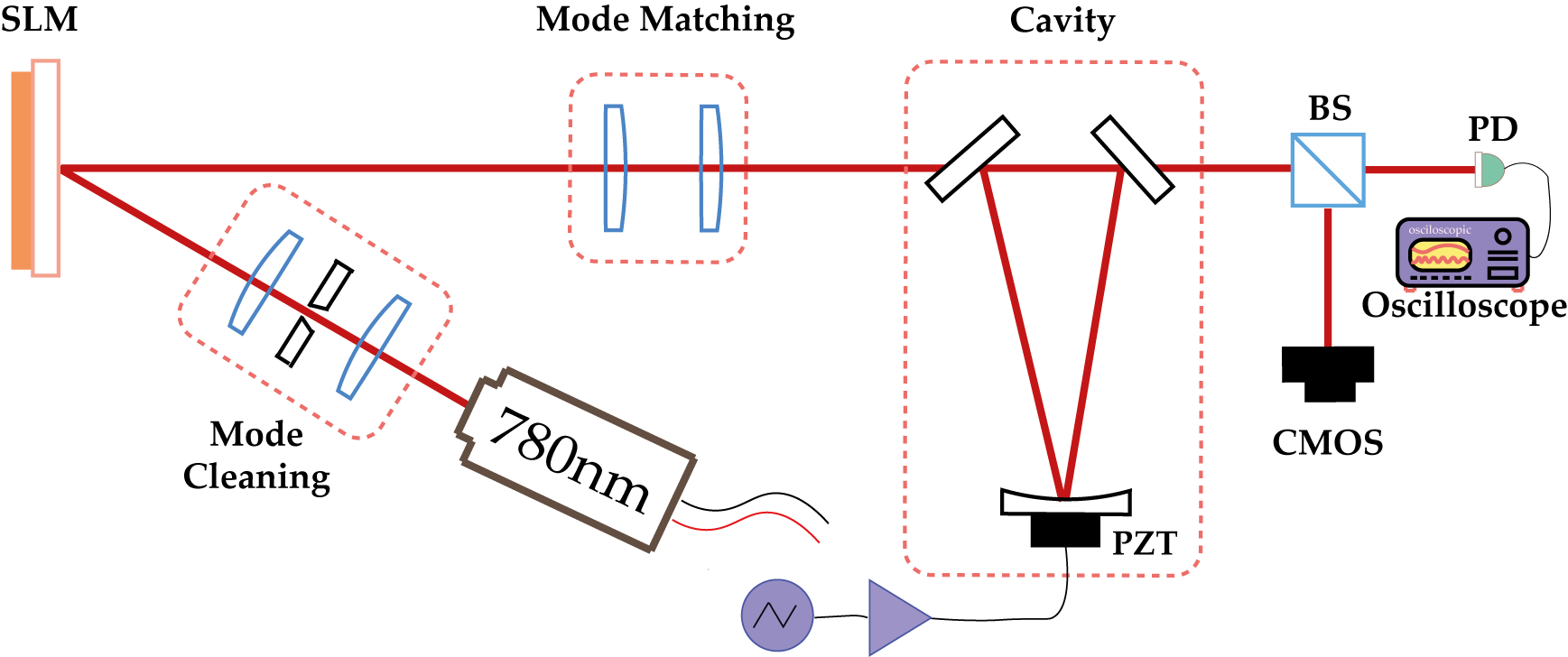}
   \caption{Experimental setup. Mode cleaning: a pair of lens in telescopic configuration with a 10-$\mu$m pinhole at the focal point; SLM: Spatial Light Modulator; Mode matching: a pair of lenses acting on the incoming beam to match the cavity waist; PZT: piezoelectric actuator; BS: beam-splitter; PD: (bulk) photodetector; CMOS: pixelated sensor consisting of a repurposed inexpensive webcam with its lens removed. Oscilloscope: used to observe the resonance peaks while the cavity length is periodically swept by the PZT.}
   \label{fig:setup}
\end{figure}

In this section, we describe the experimental setup, illustrated in Fig. \ref{fig:setup}), designed to show that a triangular cavity is capable of detecting and distinguishing between vortex and pseudovortex beams. The distinction is evaluated by inspecting their resonance peaks, which should reproduce the decompositions (\ref{eq:V-decomposition}) and (\ref{eq:PV-decomposition}), respectively.

%%%%%%%%%%%%%%%%%%%%%%%%%%%%%%%%%%%%%%%%%%%%%%%
% FIG
%%%%%%%%%%%%%%%%%%%%%%%%%%%%%%%%%%%%%%%%%%%%%%%
%%%%%%%%%%%%%%%%%%%%%%%%%%%%%%%%%%%%%%%%%%%%%%%%

\begin{figure}[H]
    \centering
    \includegraphics[width=0.9\linewidth]{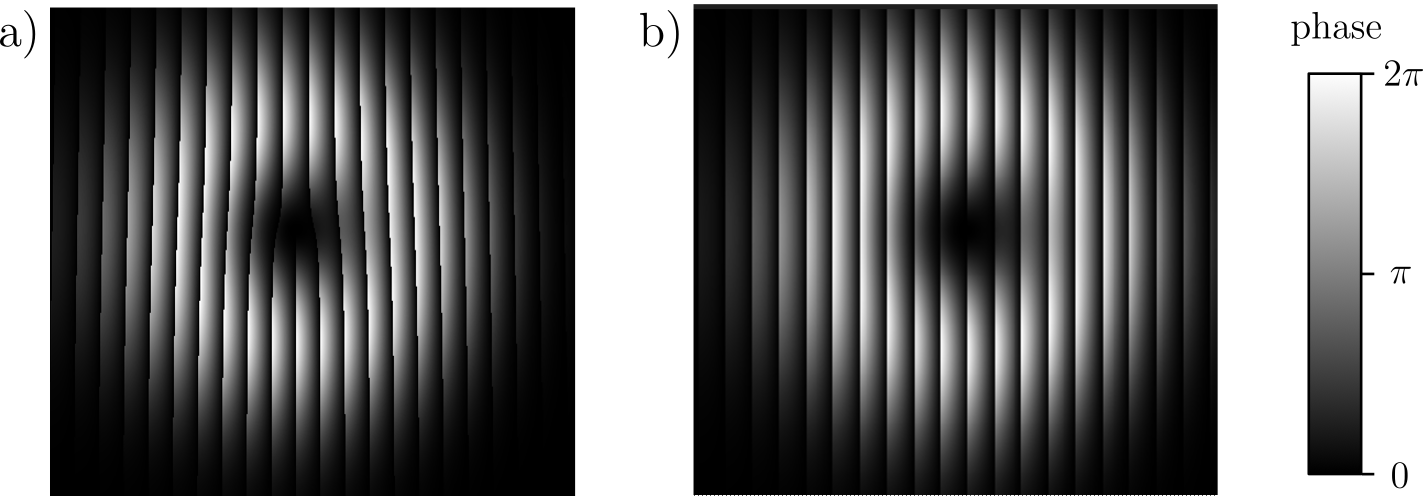}
    \caption{Hologram phase masks used for the generation of a) vortex and b) pseudovortex beams.}
    \label{fig:hologram_mask}
\end{figure}

The laser source used in this experiment is an external cavity diode laser operating in continuous wave mode at a wavelength of 780 nm. After passing through a mode-cleaning and collimation system, the beam is directed to a Spatial Light Modulator (SLM). The SLM is a phase modulator consisting of a liquid crystal screen used here to create programmed holographic masks applied directly to the beam transverse profile. To generate the optical vortex, we used a forked grating mask \cite{stoyanov2015far}, combined with amplitude modulation \cite{rosales2017shape} that reproduces the doughnut-like profile of the $LG_{10}$ mode (Fig. \ref{fig:hologram_mask}a)). For the pseudovortex, we combined a simple blazed grating phase pattern (without azimuthal phase, and thus no fork pattern) with the doughnut-like amplitude modulation  (Fig. \ref{fig:hologram_mask}b)). The amplitude modulation defines a beam waist that, together with the mode matching lenses, ensures that the beam reaches the cavity with the correct waist, which is 131 $\mu$m in our case.

%%%%%%%%%%%%%%%%%%%%%%%%%%%%%%%%%%%%%%%%%%%%%%%
% FIG
%%%%%%%%%%%%%%%%%%%%%%%%%%%%%%%%%%%%%%%%%%%%%%%
%%%%%%%%%%%%%%%%%%%%%%%%%%%%%%%%%%%%%%%%%%%%%%%

The cavity is composed of two identical partially reflective flat mirrors and a high-reflectance concave mirror. A piezoelectric actuator is positioned behind the concave mirror to scan the cavity length at  $\approx 10Hz$. The transmitted beams are sent to the photodetector, which is connected to an oscilloscope, where the resonance peaks are observed. A CMOS sensor (a repurposed inexpensive webcam with its lens removed) is placed at the other output of the beam splitter to record the intensity profile of each individual peak.

%%%%%%%%%%%%%%%%%%%%%%%%%%%%%%%%%%%%%%%%%%%%%%%
%  SECTION
%%%%%%%%%%%%%%%%%%%%%%%%%%%%%%%%%%%%%%%%%%%%%%%
\section{Results}
\label{sec:results}

\begin{figure}[h!]
\centering
   \includegraphics[width=\columnwidth]{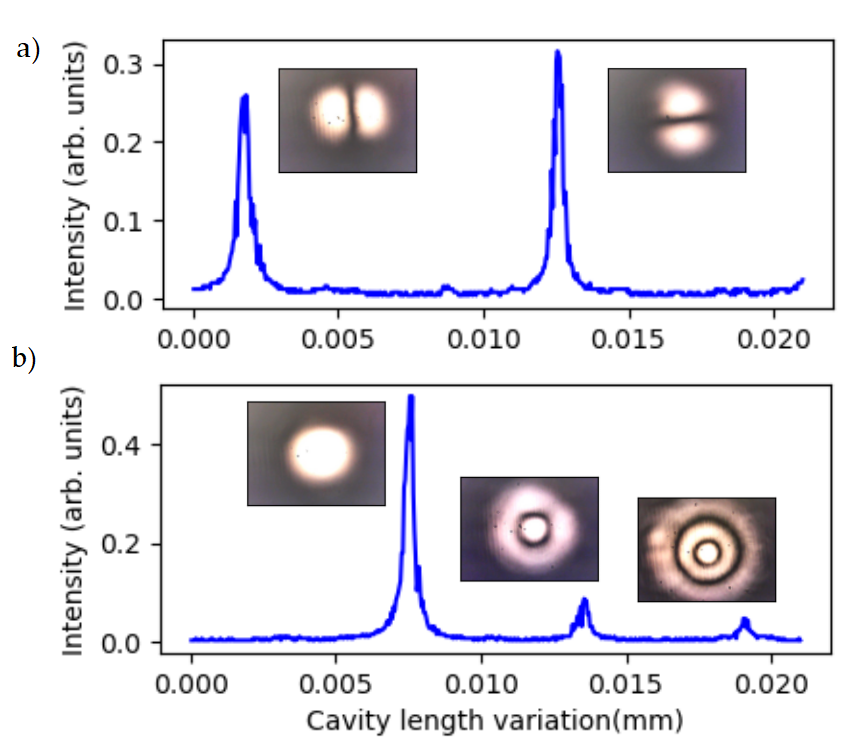}
   \caption{Resonances peaks of a) vortex $V(r,\phi)$ and b) pseudovortex $PV(r,\phi)$. The relative height of the peaks are sufficient to discriminate the vortex and the pseudovortex. The absolute height of each peak is not relevant for discrimination and depends on the overall optical power sent to the cavity.}
   \label{fig:peaks}
\end{figure}

Figure \ref{fig:peaks}) compares the resonance peaks of the vortex and the pseudovortex beams ($V(r, \varphi, z)$ and $PV(r,\varphi,z)$) generated by the triangular optical cavity. Each subfigure displays the transmitted intensity at the cavity output as a function of cavity length.

To understand these results, let us first consider the resonance of a mode with arbitrary order $N$. For constructive interference within the cavity, the total phase accumulated over a round-trip must be a multiple of $2\pi$. Other than the plane-wave phase $2kL$, where $2L$ is the total cavity length, two main factors determine the resonance lengths of this mode: 
\begin{itemize}
    \item the Gouy phase $\Delta \varphi = 2 (N+1) \tan^{-1}\left(\frac{L}{z_0}\right)$ (where $z_0$ is the Rayleigh length); and
    \item the phase introduced by reflection on cavity mirrors.
\end{itemize}
If only the effect of the Gouy phase were to be considered in a triangular cavity, modes of the same order would resonate at the same cavity lengths. However, we must also account for the effect of the odd number of cavity mirrors. When a mode reflects on a mirror at nonnormal incidence, modes that are antisymmetric with respect to a horizontal flip acquire an extra phase of $\pi$ in relation to the symmetric modes, due to the inversion of the horizontal axis in the reflection transformation \cite{rodrigues2024resonance,Santos2021}. As a result, the $HG_{01}$ and $HG_{10}$ modes resonate at different cavity lengths in a triangular cavity, although they are of the same order. 

This explains the resonance pattern of the vortex beam in Fig. \ref{fig:peaks}a), which contains two peaks of approximately the same height, as expected from decomposition (\ref{eq:V-decomposition}). The modes corresponding to each peak were captured on camera when the cavity length was slowly swept and are displayed in the insets.

Figure \ref{fig:peaks}b) shows the resonance pattern of the pseudovortex $PV(r,\varphi,z)$. The insets demonstrate its decomposition in the Laguerre-Gauss basis, in accordance with Eq. (\ref{eq:PV-decomposition}). The heights of the peaks express the relative contribution of each mode to the total intensity and should be proportional to the square of the coefficient of each mode in the decomposition. The predicted ratio between the heights of the $LG_{00}$ and $LG_{01}$ peaks is according to Eq. (\ref{eq:PV-decomposition}), exactly $\left(\frac{\sqrt{\pi}}{2}/\frac{\sqrt{\pi}}{4}\right)^2=4$. Experimentally, we obtain a ratio of approximately 5, showing a good enough agreement that enables one to undoubtedly distinguish between the vortex and pseudovortex patterns produced by the triangular cavity. Error sources that could explain the slight ratio discrepancy include cavity-length and laser-intensity instability, and SLM calibration imperfections leading to limited mode purity.

%%%%%%%%%%%%%%%%%%%%%%%%%%%%%%%%%%%%%%%%%%%%%%%
% FIG
%%%%%%%%%%%%%%%%%%%%%%%%%%%%%%%%%%%%%%%%%%%%%%%
%%%%%%%%%%%%%%%%%%%%%%%%%%%%%%%%%%%%%%%%%%%%%%%%

\section{Discussion}
\label{sec:discussion}

This interferometric technique may also be employed to distinguish vortices with higher topological charges (Eq. \ref{eq:LG_lp} with increasing $\ell$) from their corresponding pseudovortices (Eq. \ref{eq:LG_lp} without the azimuthal phase factor $\e^{\ii\ell\phi}$). 

In fact, independently of how large the index $\ell$ is, a laguerre-gaussian beam $LG_{\ell 0}$ can always be decomposed into a symmetric and an antisymmetric component (with respect to a horizintal flip). This leads to the observation of exactly two resonance peaks (one for each component) separated by a phase of $\pi$, which is equivalent to half a wavelength in terms of the round-trip length of the cavity.

\begin{figure}[h!]
\centering
   \includegraphics[width=\columnwidth]{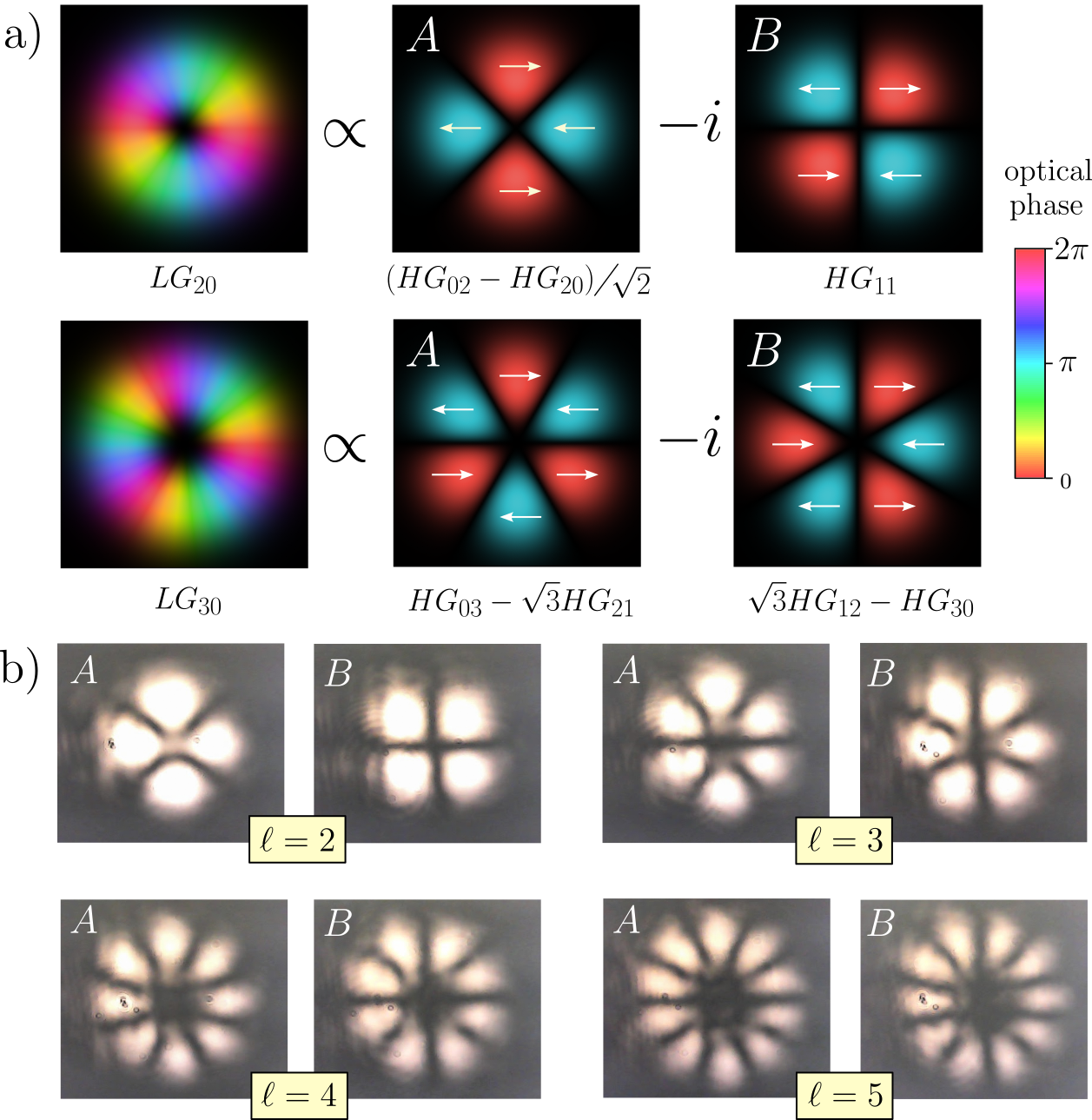}
   \caption{a) Decomposition of horizontally polarized OAM modes with $\ell = 2$ (top row) and $\ell = 3$ (bottom row) into symmetric and antisymmetric components under horizontal reflection (i.e., even and odd parity with respect to the vertical axis). Component $A$ has an antisymmetric electric vector field, while component $B$ is symmetric --mirroring $B$ across the vertical axis does not alter the orientation of the electric field vectors (indicated by white arrows). For vertically polarized beams, the symmetry of components $A$ and $B$ is reversed: $A$ becomes symmetric and $B$ antisymmetric. b) Experimental intensity profiles of ``camomile'' modes measured at the output of the triangular cavity. Each image pair corresponds to an input $LG_{\ell 0}$ mode; components $A$ and $B$ are associated with distinct resonance peaks within the cavity.}
   \label{fig:camomile}
\end{figure}

By way of example, Fig. \ref{fig:camomile}a) illustrates the decomposition of horizontally polarized vortex beams $LG_{20}$ and $LG_{30}$ into their symmetric and antisymmetric components, resulting in patterns resembling a ``camomile'', as termed by Khonina \emph{et al.} \cite{Khonina2004}.

In contrast, a pseudovortex, as previously mentioned, is a radially symmetric mode that does not exhibit orbital angular momentum. Consequently, it decomposes into a series of Laguerre-Gauss modes with $\ell = 0$ and varying radial indices $p$ (and, therefore, varying mode order $N=2p$). Table \ref{tab:intensities} presents the intensity contributions of $LG_{0p}$ modes to pseudovortices as $|\ell|$ increases.

All things considered, an $LG_{\ell 0}$ vortex consistently exhibits a pair of resonance peaks separated by half a free spectral range, allowing it to be distinguished from its corresponding pseudovortex, whose peak pattern varies according to Table \ref{tab:intensities}. The same reasoning applies to the case of $p > 0$: any $LG_{\ell p}$ vortex (of order $N=|\ell|+2p$) can be decomposed into $N+1$ $HG_{mn}$ modes (of the same order $N=m+n$); as modes of the same order acquire the same Gouy phase over a cavity round-trip, the only phase distinguishing them is the one due to parity; thus, symmetric and antisymmetric modes will form two separate groups of modes, each resonating at a specific cavity length. In contrast, $LG_{\ell p}$ pseudovortices decompose into a diverse set of modes with varying orders.

%%%%%%%%%%%%%%%%%%%%%%%%%%%%%%%%%%%%%%%%%%%%%%%
% FIG
%%%%%%%%%%%%%%%%%%%%%%%%%%%%%%%%%%%%%%%%%%%%%%%
%%%%%%%%%%%%%%%%%%%%%%%%%%%%%%%%%%%%%%%%%%%%%%%%

\begin{table}[]
\begin{tabularx}{\columnwidth}{ >{\centering}X | >{\centering}X  >{\centering}X  >{\centering}X  >{\centering\arraybackslash}X }
$|\ell|$ & $LG_{00}$ & $LG_{01}$ & $LG_{02}$ & $\sum_{p>2}LG_{0p}$ \\ [1ex] \hline
1      & 78.5      & 19.6      & 1.2       & 0.6                 \\
2      & 50.0      & 50.0      & 0.0       & 0.0                 \\
3      & 29.4      & 66.3      & 4.1       & 0.1                 \\
4      & 16,7      & 66.7      & 16.7      & 0.0                 \\
5      & 9.2       & 57.5      & 32.3      & 0.9                 \\
6      & 5.0       & 45.0      & 45.0      & 5.0                 \\
7      & 2.7       & 32.9      & 51.3      & 13.0                \\
8      & 1.4       & 22.9      & 51.4      & 24.3                \\
9      & 0.8       & 15.3      & 46.8      & 37.1                \\
10     & 0.4       & 9.9       & 39.7      & 50.0                \\ 
\end{tabularx}
\caption{Mode decomposition of pseudovortices with different values of $\ell$ on the Laguerre-Gauss basis. Numbers are the result of theoretical calculations and indicate the percentage of the mode intensity, that is, $100\times$ the square of the corresponding coefficient in mode decomposition. Each line sums 1 (up to truncation imprecisions).}
\label{tab:intensities}
\end{table}

In the above, we investigated the differences between the interaction of vortices and pseudovortices with a triangular optical cavity. We now turn to possible applications.

First, from Fig. \ref{fig:camomile}), it becomes clear that the triangular cavity can be used to separate ``camomile'' modes: by putting one mode in resonance, the other will be reflected by the input mirror. This property could be applied in binary communication protocols that exploit superpositions of such modes, with the cavity playing a key role in reading and manipulating these superpositions.

%%%%%%%%%%%%%%%%%%%%%%%%%%%%%%%%%%%%%%%%%%%%%%%
% FIG
%%%%%%%%%%%%%%%%%%%%%%%%%%%%%%%%%%%%%%%%%%%%%%%
\begin{figure}[h!]
\centering
   \includegraphics[width=\columnwidth]{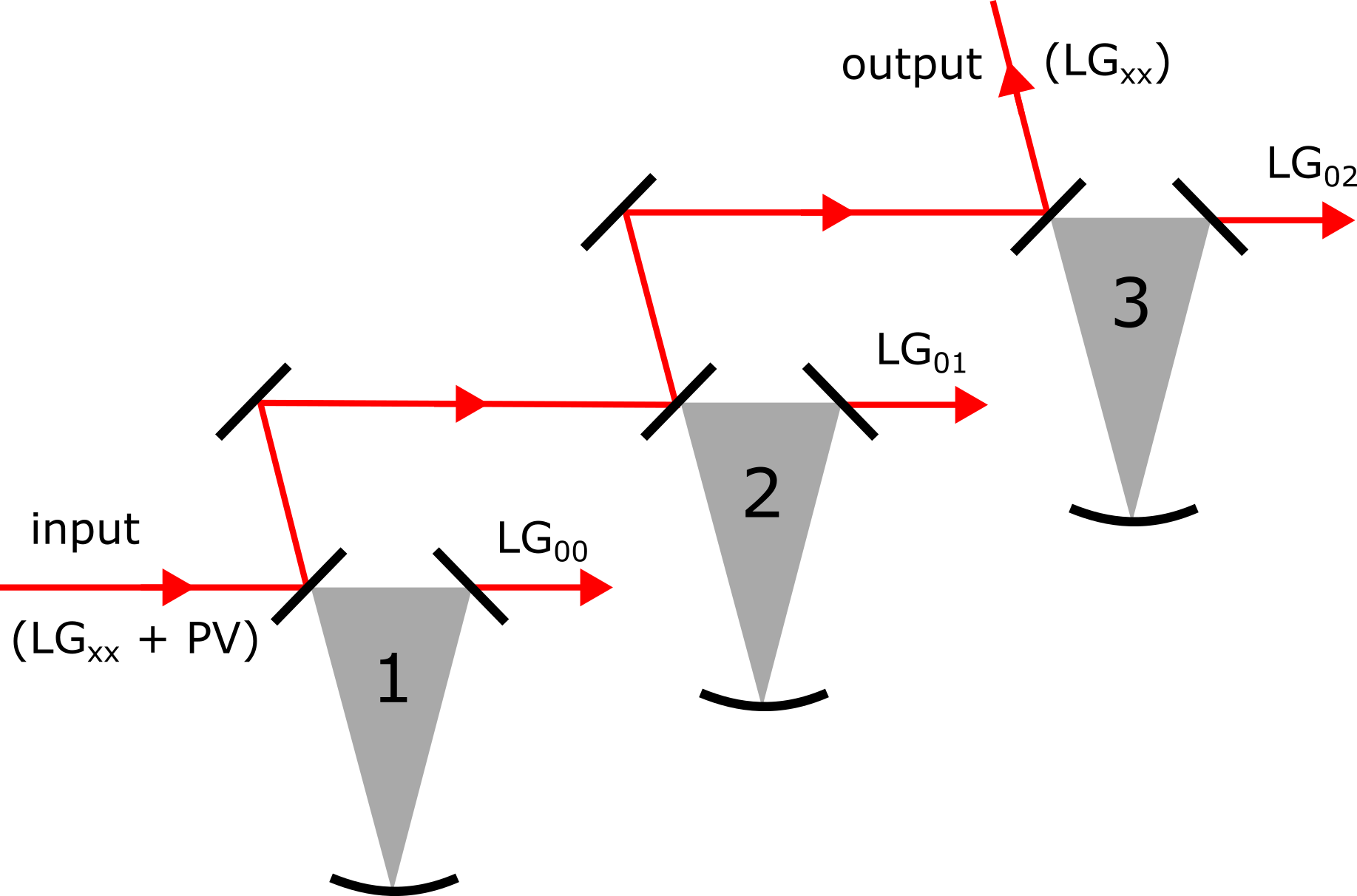}
   \caption{Proposed scheme for filtering out a pseudovortex (PV) from a Laguerre-Gaussian vortex beam.}
   \label{fig:cavities}
\end{figure}
%%%%%%%%%%%%%%%%%%%%%%%%%%%%%%%%%%%%%%%%%%%%%%%%

Second, the triangular cavity setup can also be used to separate modes with specific topological charges from pseudovortices (modes without OAM). Figure \ref{fig:cavities}) illustrates a system consisting of three triangular cavities. The input beam is first directed to cavity 1, which is tuned to resonate with the $LG_{00}$ mode. As a result, this mode is transmitted, while all other non-resonant modes are reflected and sent to cavity 2. Cavity 2 is tuned to resonate with the $LG_{01}$ mode, allowing it to pass through while further filtering out $LG_{00}$ and $LG_{01}$ from the reflected light. This reflected beam is then directed to cavity 3, which resonates with the $LG_{02}$ mode. The $LG_{02}$ mode is transmitted, leaving the remaining reflected light free of $LG_{00}$, $LG_{01}$, and $LG_{02}$. Numerical calculations (see Table \ref{tab:intensities}) show that, after filtering these three components, the output light is a vortex mode with a fidelity exceeding 99\% for Laguerre-Gaussian modes with topological charges $|\ell| = 1,\ 3$, or 5. The process should also work for $|\ell| = 6$ or 7, with fidelity exceeding 85\%.

The process does not work for $|\ell| = 2$ or 4, however, because the modes $LG_{20}$ and $LG_{40}$ have the same order as the modes $LG_{01}$ and $LG_{02}$, respectively (since $N = |\ell| + 2p$). As a result, attempting to extract the mode $LG_{01}$ from a superposition containing the vortex $LG_{20}$, for example, would inadvertently remove the symmetric component of this vortex (or the antisymmetric component, depending on polarization (see Fig. \ref{fig:camomile})). This happens because the symmetric (or antisymmetric) component of $LG_{20}$ and the mode $LG_{01}$ resonate together, that is, are transmitted for the same cavity length. Despite this limitation, the scheme works for a range of $\ell$ values (both positive and negative) and may still be useful. It is worth noting that this technical constraint cited above is inherent to optical cavities, which only discerns modes based on their order and (if the number of mirrors is odd) their parity.

Although the concept illustrated in Fig. \ref{fig:cavities}) is straightforward, its practical implementation is left for future work. We note that this implementation involves positioning mode-matching lenses between the cavities to ensure that the beam is focused with the appropriate waist exactly halfway between the mirrors.

There are several methods for characterizing a light beam based on its Laguerre-Gaussian (LG) or Hermite-Gaussian (HG) mode components (see, for example, Refs. \cite{Hickmann2010, Kotlyar2017, Volyar2019, Kumar2022}. These include techniques that rely on detecting the light and analyzing its intensity patterns. Another class of approaches involves sorting LG and HG optical modes -- where each mode component is mapped onto a different degree of freedom, such as the direction of propagation -- eliminating the need for direct intensity measurements or photon counting (see Refs. \cite{Berkhout2010, Mirhosseini2013, Wen2020}). Our scheme belongs to this second category of mode sorters but introduces a key distinction: we exploit the resonance conditions of a sequence of optical cavities to progressively filter out a pseudovortex mode -- a specific superposition in the LG basis -- while keeping a vortex mode at the output. To the best of our knowledge, no previous schemes have demonstrated this type of mode sorting.

%%%%%%%%%%%%%%%%%%%%%%%%%%%%%%%%%%%%%%%%%%%%%%%
%  SECTION
%%%%%%%%%%%%%%%%%%%%%%%%%%%%%%%%%%%%%%%%%%%%%%%
\section{Conclusion}
\label{sec:conc}
In summary, we have theoretically and experimentally demonstrated that a triangular cavity can effectively discriminate between vortex and pseudovortex beams. This is attributed to the inherent differences in their wavefront structures, in conjunction with the geometry of the triangular cavity, leading to distinct resonance peaks. These peaks can be identified and selectively tuned by adjusting the cavity length. The experimental data aligns closely with our theoretical predictions.

From an application perspective, we propose that such a cavity could serve as a sorter for ``camomile'' modes, which are the symmetric/antisymmetric components of a pure Laguerre-Gauss vortex. Additionally, a scheme using three triangular cavities could be employed to filter out pseudovortices in superposition with real vortices. These findings represent a step forward in using triangular cavities as efficient mode-sorting elements, with potential applications in the development of optical devices for communication systems.

\section*{Appendix}

In this appendix, we calculate the coefficients $a_{0p}$ of the pseudovortex LG-decomposition. Eq.~(\ref{eq:LG_lp}) for $\ell=0$ yields
\begin{equation}
    LG_{0p}(r,\phi) = \frac{A_{0p}}{w_0}\, L_p\!\left(\frac{2r^2}{w_0^2}\right)\e^{-r^2/w_0^2}\,,
\end{equation}
where $L_p$ are the Laguerre polynomials, which are equal to the \textit{generalized} Laguerre polynomials $L^\alpha_p$ for $\alpha=0$. Using the fact that $A_{0p}=\sqrt{2/\pi}$ and by performing a simple change of variables, $u=\sqrt{2}\,r/w_0$, eq.~(\ref{eq:a_lp}) simplifies to
\begin{equation}
    a_{0p} = \int_{-\infty}^{+\infty} \!du\,u^2\,L_p(u^2)\,\e^{-u^2}\,.
    \label{eq:a_0p}
\end{equation}

Now, by the definition of Laguerre Polynomials
\begin{equation}
    L_p(u^2)=\sum_{n=0}^p \frac{(-1)^n}{n!}\binom{p}{n}\,u^{2n}
\end{equation}
and using the fact that
\begin{equation}
    \int_{-\infty}^{+\infty} \!du\,u^{2n+2},\e^{-u^2}=\frac{(2n+1)!!}{2^{n+1}}\sqrt{\pi}\,,
\end{equation}
Eq.~(\ref{eq:a_0p}) leads to
\begin{align}
    a_{0p} &= \frac{\sqrt{\pi}}{2} \sum_{n=0}^p \frac{(-1)^n}{2^n}\,\frac{(2n+1)!!}{n!}\,\binom{p}{n} \nonumber\\
     &= -\frac{\Gamma(p-1/2)}{4\Gamma(p+1)}\,.
\end{align}
From the following properties of the Gamma function:
\begin{equation}
    \Gamma(0) = 1,\ \Gamma(-1/2)=-2\sqrt{\pi},\ \Gamma(z+1)=z\Gamma(z), 
\end{equation}
we recover the coefficients explicitly shown in Eq.~(\ref{eq:PV-decomposition}).

\begin{acknowledgments}
We thank Prof. Marcelo Martinelli for providing us with the laser source. We also acknowledge funding from the Brazilian agencies: Conselho Nacional de Desenvolvimento Tecnol\'ogico (CNPq), Funda\c c\~{a}o de Amparo \`{a} Pesquisa e Inova\c{c}\~{a}o do Estado de Santa Catarina (FAPESC), Coordena\c c\~{a}o de Aperfei\c coamento de Pessoal de N\'ivel Superior - Brasil (CAPES) and Instituto Nacional de Ci\^encia e Tecnologia de Informa\c c\~ao Qu\^antica (INCT/IQ 465469/2014-0).
\end{acknowledgments}

\section*{CRediT authorship contribution statement}

\textbf{L. Marques Fagundes}: conducted the experiment, analyzed the results, wrote and reviewed the manuscript. \textbf{P. H. Souto Ribeiro}: provided financial support, wrote and reviewed the manuscript. \textbf{R. Medeiros de Araújo}: conceived the experiment, carried out theoretical deductions, wrote and reviewed the manuscript.

\bibliography{bibliopseudovortex}

\end{document}